\documentclass[prb,preprint]{revtex4}
\usepackage[T1]{fontenc}
\usepackage[latin9]{inputenc}
\usepackage{graphicx}

\makeatletter
\@ifundefined{textcolor}{}
{%
 \definecolor{BLACK}{gray}{0}
 \definecolor{WHITE}{gray}{1}
 \definecolor{RED}{rgb}{1,0,0}
 \definecolor{GREEN}{rgb}{0,1,0}
 \definecolor{BLUE}{rgb}{0,0,1}
 \definecolor{CYAN}{cmyk}{1,0,0,0}
 \definecolor{MAGENTA}{cmyk}{0,1,0,0}
 \definecolor{YELLOW}{cmyk}{0,0,1,0}
}

\makeatother

\begin{document}

\title{Magnetic properties of carbon nanodisk and nanocone powders }

\author{Jozef \v{C}ern\'{a}k}

\affiliation{Institute of Physics, P. J. \v{S}af\'{a}rik University in Ko\v{s}ice,
Jesenn\'{a} 5, SK-04000 Ko\v{s}ice, Slovak Republic}

\author{Geir Helgesen}

\affiliation{Institute for Energy Technology, Physics Department, NO-2007 Kjeller,
Norway }

\author{Jozef Kov\'{a}\v{c}}

\affiliation{Laboratory of nanomaterials and applied magnetisms, Institute of
Physics, Slovak Academy of Sciences, Watsonova 47, SK-04000 Ko\v{s}ice,
Slovak Republic}

\author{Arne T. Skjeltorp}

\affiliation{Institute for Energy Technology, Physics Department, NO-2007 Kjeller,
Norway }

\author{Josef Voltr}

\affiliation{Faculty of Nuclear Sciences and Physical Engineering, Czech Technical
University in Prague, B\v{r}ehov\'{a} 7, CZ-115 19 Praha 1, Czech
Republic}

\author{Erik \v{C}i\v{z}m\'{a}r}

\affiliation{Centre of Low Temperature Physics, P. J. \v{S}af\'{a}rik University
in Ko\v{s}ice, SK-04000 Ko\v{s}ice, Slovak Republic}
\begin{abstract}
We have investigated the magnetic properties of carbon powders which
consist of nanodisks, nanocones, and a small fraction of carbon-black
particles. Magnetization measurements were carried out using a superconducting
quantum interference device in magnetic fields $-5<\mu_{0}H<5\:\mathrm{T}$
for temperatures in the range $2\leq T<350\:\mathrm{K}$. Measurements
of the magnetization $M$ versus temperature $T$ and magnetic field
$\mu_{0}H$ for these carbon samples show diamagnetism and paramagetism
with an additional ferromagnetic contribution. The ferromagnetic magnetization
is in agreement with the calculated magnetization from Fe impurities
as determined by the particle-induced x-ray emission method ($<75\:\mu\mathrm{g/g}$).
Magnetization measurements in weak magnetic fields show thermal hysteresis,
and for strong fields the magnetization $M$ decreases as $M\sim aT^{-\alpha}$
with $\alpha<1$, which is slower than the Curie law ($\alpha=1$),
when the temperature increases. The magnetization $M$ versus magnetic
field $\mu_{0}H$ shows paramagnetic free-spin $S=\frac{1}{2}$ and
$\frac{3}{2}$ behaviors for temperatures $T=2\:\mathrm{K}$ and $15\leq T\leq50\:\mathrm{K}$,
respectively. A tendency for localization of electrons was found by
electron spin resonance when the temperature $T$ decreases ($2<T<40\:\mathrm{K}$).
The magnetic properties in these carbon cone and disk powder samples
are more complex than a free-spin model predicts, which is apparently
valid only for the temperature $T=2\:\mathrm{K}$.
\end{abstract}

\pacs{75.20.Ck}

\maketitle

\section{Introduction}

Carbon atoms can bind via $\sigma$ and $\pi$ bonds when forming
a molecule. The number and nature of the bonds determine the geometry
and properties of carbon allotropes. \cite{Hennrich} Elemental carbon
naturally forms three well-known allotropes: graphite, diamond, and
carbon black. In the past, new carbon allotropes have been synthesized:
fullerenes, carbon nanotubes, \cite{Hirch} and graphene. \cite{Castro_Neto}
Two recently published papers \cite{Sheng,Hirch} illustrate an effort
to propose and design new carbon allotropes. The structure of carbon
nanocones \cite{Krishnan} containing a small number of pentagons
in a graphene-like layer of hexagons is the reason why nanocones often
have been considered to be a specific kind of fullerene. \cite{Gogotsi}
However, the nanocones and nanodisks differ from fullerenes in shape
\cite{Krishnan} and wall thickness which may be from a few up to
several tens of graphene layers. These differences could be a reason
for their different properties relative to the fullerenes.

Elemental-carbon-based materials show a diversity of electronic properties
metallic, semiconducting, or dielectric, \cite{Castro_Neto} but they
are commonly classified as semiconductors. \cite{Makarova} For example,
a single graphene layer or stack of a few graphene layers can display
Dirac-like electron excitations which result in unusual spectroscopic
and transport properties. \cite{Castro_Neto} The magnetic properties
of graphite are diamagnetic due to the delocalized $\pi$-band electrons.
On the other hand, diamond displays paramagnetic magnetization as
a consequence of localized electrons. Flow of currents around the
carbon rings of graphite in response to an applied magnetic field
has been used to explain the differences between the susceptibility
of graphite and that of diamond found in experiments. \cite{Haddon}
The fullerenes can exhibit both diamagnetic and paramagnetic ring
currents which lead to subtle effects in the magnetic properties of
these molecules and provide evidence for the existence of $\pi$ electrons
mobile in three dimensions. \cite{Haddon}

The early reports on possible ferromagneticlike behavior in carbon
structures were not generally accepted by the scientific community.
\cite{Esquinazi_2007} It was initially assumed that ferromagnetic
behavior results from residual amounts of ferromagnetic impurities
(Fe, Ni, or Co) in the carbon samples. A systematic study performed
by H\"{o}hne \textit{et al}.\cite{Hohne_2008} did not show any influence
of iron atoms on the ferromagnetic properties of highly oriented pyrolitic
graphite (HOPG) up to Fe impurity densities of $\sim4000\;\mu\mathrm{g/g}$,
and this supported the initial assumption that uniformly distributed
iron up to $100$ ppm cannot trigger ferromagnetic order. \cite{Makarova}
The reason is that uniformly distributed residual magnetic impurities
can be considered to be noninteracting magnetic moments. \cite{Kopelevich}
However, recently Nair \textit{et al}. \cite{Nair} found Fe microparticles
which were attached to the surface of HOPG samples. These bigger Fe
microparticles behave in a quite different manner from uniformly distributed
Fe nanoparticles and could contribute in a ferromagnetic way to the
sample magnetization.

Carbon nanofoams \cite{Rode} and nanodiamond powders \cite{Levin}
have the common feature that their magnetizations $M$ vs temperature
$T$ show paramagnetic behavior in a wide temperature range. However,
the reasons for their paramagnetism are different. The paramagnetic
behavior of the carbon nanofoams \cite{Rode} is considered to be
a consequence of a metal-insulator-like transition which can take
place \cite{Hellman} for temperatures $T<30\:\mathrm{K}$. In the
case of nanodiamond powder, \cite{Levin} the paramagnetic magnetization
is associated with localized electrons in a wide temperature range.
Sepioni \emph{et al.} \cite{Sepioni} have investigated graphene nanocrystals
of size $10$ to $50\:\mathrm{nm}$ and thickness of one or two graphene
layers. They observed a strong diamagnetic behavior and found only
a weak paramagnetism caused by unpaired electrons at edges for low
temperatures $2\leq T\leq50\:\mathrm{K}$. Spemann \textit{et al}.
\cite{Spermann} reported on the ferromagnetic behavior of impurity-
free regions of a $\mathrm{C}_{60}$ polymer. For the $\mathrm{Rh}-\mathrm{C}_{60}$
polymerized phase, Boukhvalov \emph{et al.} \cite{Boukhvalov} concluded
that rhombohedral distortion of $\mathrm{C}_{60}$ itself cannot induce
magnetic ordering in molecular carbon. \v{C}ervenka \emph{et al.,}
\cite{Cervenka} using superconducting quantum interference device
(SQUID) magnetization measurements at temperatures $T=5$ and $300\:\mathrm{K}$
and scanning tunneling microscopy (STM), demonstrated both diamagnetism
and ferromagnetic order at room temperates in bulk HOPG caused by
two-dimensional (2D) planes of magnetized grain boundaries propagating
though the sample. However, the existence of ferromagnetic order in
bulk HOPG samples is not conclusively confirmed. \cite{Martinez,Nair}

Gonz\'{a}lez \emph{et al.} \cite{Gonzalez} theoretically investigated
electron-electron interaction in graphene layers. They found that
topological disorder enhances the density of states and can lead to
instabilities in conductivity or magnetic properties. Park \emph{et
al. }\cite{Park} applied \emph{ab initio} spin density functional
theory to demonstrate a net magnetic moment in the building block
of schwarzite. They expected that in aromatic systems with negative
Gaussian curvature unpaired spins can be introduced by sterically
protected carbon radicals. The magnetic moment of a vacancy defect
has been determined as $1.12-1.53\:\mu_{B}$ (Bohr magneton) from
first-principle calculations. \cite{Yazyev} Experimental \cite{Shibayama,Esquinazi_2003}
and theoretical results \cite{Gonzalez,Park,Yazyev} support the hypothesis
that disorder in carbon allotropes is an important precondition in
order to observe paramagnetic or ferromagnetic magnetization.

The aim of this paper is to characterize the basic magnetic properties
of a carbon powder consisting of nanocones and nanodisks and discuss
its magnetic properties in comparison to other carbon allotropes.

The paper is organized as follows. In the next section, Sec. \ref{sec:Experiments},
the carbon powder samples and the experimental methods are described.
The results of magnetic measurements are presented in Sec. \ref{sec:Results}
and discussed in Sec. \ref{sec:Discussion}. Conclusions are given
in Sec. \ref{sec:Conclusions}.

\section{\label{sec:Experiments}Experiments}

\subsection{\label{sub:Samples}Carbon powder samples}

The graphitic carbon powder was produced by the so-called pyrolytic
Kvaerner ``carbon-black and hydrogen process'' \cite{Hugdahl}.
The powder consists of flat carbon nanodisks, open-ended carbon cones,
and a smaller amount of carbon black. \cite{Krishnan,Garberg,Heiberg-Andersen_2011}

The carbon disks and cones exhibit a wide range of diameters ($500-4000\:\mathrm{nm}$)
and their wall thickness is mainly $10-30\:\mathrm{nm}$ but particles
with thickness in the range $5-70\:\mathrm{nm}$ can be found. The
electron diffraction patterns of the nanodisks consist of concentric
continuous rings with distinct spots with six-fold rotational symmetry.
These results led to the conclusion that the nanodisks are multilayer
carbon structures with a graphitic core and outer non-crystalline
layers, \cite{Garberg} which was also supported by scanning electron
microscopy (SEM) and transmission electron microscopy (TEM) images.
TEM micrographs of the carbon powder showed the presence of perfect
carbon nanocones of all the five possible apex angles \cite{Krishnan}
$\alpha=112.9^{\circ}$, $83.6^{\circ}$, $60.0^{\circ}$, $38.9^{\circ}$,
and $19.2^{\circ}$ corresponding to $n=1-5$ carbon pentagonal rings
near the cone tip. $n=0$ corresponds to the flat disks. Later, electron
diffraction analysis of the nanocones showed that they are similar
to the disks with a graphitic core surrounded by amorphous outer carbon
layers.\cite{Naess} Some of the disks and the $112.9^{\circ}$ apex-angle
cones showed six-fold and five-fold faceting, respectively, along
their edges. The thickness of the crystalline core was estimated to
be only $10-30\%$ of the disk thickness. These cones are different
from the conical graphite crystals reported by Gogotsi \textit{et
al.} \cite{Gogotsi-2002} and carbon nanohorns. \cite{Iijima-1999}

\begin{figure}
\includegraphics[width=8.5cm]{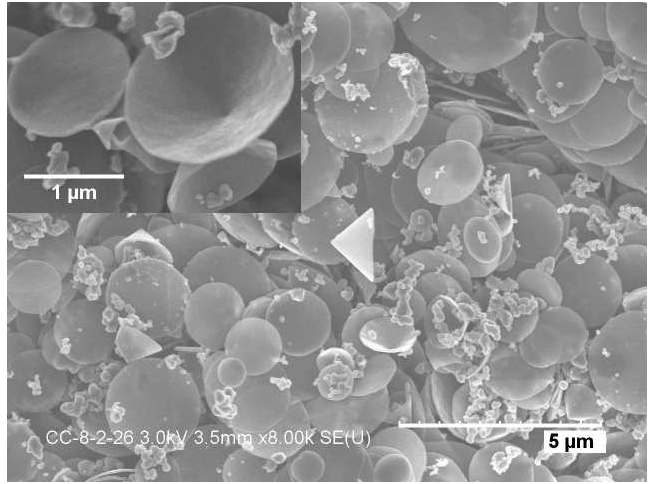}

\caption{Scanning electron microscopy micrographs of the carbon powder. The
inset shows details of the cones. }
\end{figure}

The investigated sample shows disorder on at least two length scales;
on the nanometer scale it is a mixture of crystalline parts, possibly
containing many dislocations, grain boundaries, and other defects,
and non-crystalline or amorphous matter. On the micrometer scale the
grainy nature of the powder will cause different packings of particles
and thus a varying material density.

\subsection{\label{sub:Experimental-methods}Experimental methods}

\subsubsection{\label{sub:Impurities-measurements}Iron impurity measurements}

It is quite common that carbon nanomaterials contain trace amounts
of Fe contamination. Therefore, it is important to determine the density
of magnetic impurities such as Fe in the carbon powder. Here, the
particle induced x-ray emission method \cite{pixe-S} was used to
determine the Fe content. The carbon powder was fixed on polycarbonate
membranes of diameter $25\:\mathrm{mm}$ with pore size of $5\:\mu\mathrm{m}$
(Cyclopore). In order to determine the mass of powder, the difference
between membrane mass without powder and with powder after deposition
was determined. The mass of powder was in the range $0.1-0.2\:\mathrm{\mathrm{m}g}$
(using a Mettler Toledo balance). Taking into account the density
of amorphous carbon $\sim2300\,\mathrm{\mathrm{k}g^{-3}}$, the thickness
of the deposit was in the range $2-4\:\mu\mathrm{m}$. A spot of diameter
$5\:\mathrm{mm}$ was irradiated by a proton beam of energy $1.8\:\mathrm{M}\mathrm{e}\mathrm{V}$
from a Van de Graaff accelerator at the ion beam facilities at the
Czech Technical University. The diameter of the ion beam was $8\mathrm{\: mm}$.
Samples were placed pependicular to the beam axis and a Si(Li) detector
was used to collect x-rays at an angle of $60^{o}$. In order to attenuate
the low- energy part of the spectra, a Mylar foil of $356\:\mu\mathrm{m}$
thickness was placed in front of the detector. Beam doses of $15-30\:\mu\mathrm{C}$
were measured using a Faraday cup behind the sample. A thin standard
of Fe (MicroMatter) was used to obtain the reference energy spectra
of Fe. Energy spectra were calculated by the GUPIXWIN software package,
\cite{GUPIX} where background spectra of the polycarbonate membrane
were subtracted. The mass of Fe was related to the mass of carbon
material. The powder samples were analyzed after magnetization measurements
in order to determine the possible Fe contamination during sample
handling and manipulation. Fe concentrations of about $60$$\pm15$
$\mu\mathrm{g/\mathrm{g}}$ were found for several of these powder
samples.

\subsubsection{Electron spin resonance}

In order to determine possible deviations from the normal electron
spin g-factor electron spin resonance (ESR) measurements were performed.
These were done using a Bruker ELEXSYS E500 X-band spectrometer working
in the temperature range $2-300\mathrm{\: K}$ on powder samples mixed
with Apiezon-N grease and attached to a Suprasil sample holder.

\subsubsection{Magnetization measurements}

Magnetization vs temperature $M(T)$ and magnetization vs magnetic
field $M(\mu_{0}H)$ measurements were carried out using a Quantum
Design dc-ac SQUID magnetic properties measurement system (MPMS) magnetometer
with a scan length of 4 cm. The samples were placed in the SQUID chamber
before magnetization measurements for $4\:\mathrm{h}$ to reduce the
concentration of oxygen in the sample. The pressure of helium gas
near the sample in the SQUID chamber was about $666\:\mathrm{Pa}$.
Zero-field-cooled (ZFC) and field-cooled (FC) protocols were used
to measure the temperature dependence of the magnetization. During
ZFC the samples were cooled to $2\:\mathrm{K}$ in zero field. Once
the temperature was stabilized, the magnetic field was applied. The
magnetic moment was measured as a function of temperature up to room
temperature $T=300\:\mathrm{K}$. For FC the samples were cooled in
the same constant field to the lowest temperature $T=2\:\mathrm{K}$
and magnetization $M(T)$ was measured. 

Magnetization vs magnetic field $M(\mu_{0}H)$ was measured for a
few, selected temperatures in the range $2\leq T\leq300\:\mathrm{K}$
in a varying magnetic field $-5\leq\mu_{0}H\leq5\:\mathrm{T}$. Several
carbon powder samples of mass $m=30-70\:\mathrm{mg}$ were encapsulated
in gelatin capsules of volume $0.13\:\mathrm{ml}$ and measured in
order to check the reproducibility of the results. Here, the results
of the study of a sample of mass $m=50\:\mathrm{mg}$ will be presented.
We have assumed that the density of the powder in the capsule is high
enough to prevent free rotation of carbon particles caused by the
magnetic field at any of the investigated temperatures. However, nano-scale
movements of particles limited by surrounding particles cannot be
excluded.

\section{\label{sec:Results}Results}

\subsection{ESR}

\begin{figure}
\includegraphics[width=8cm]{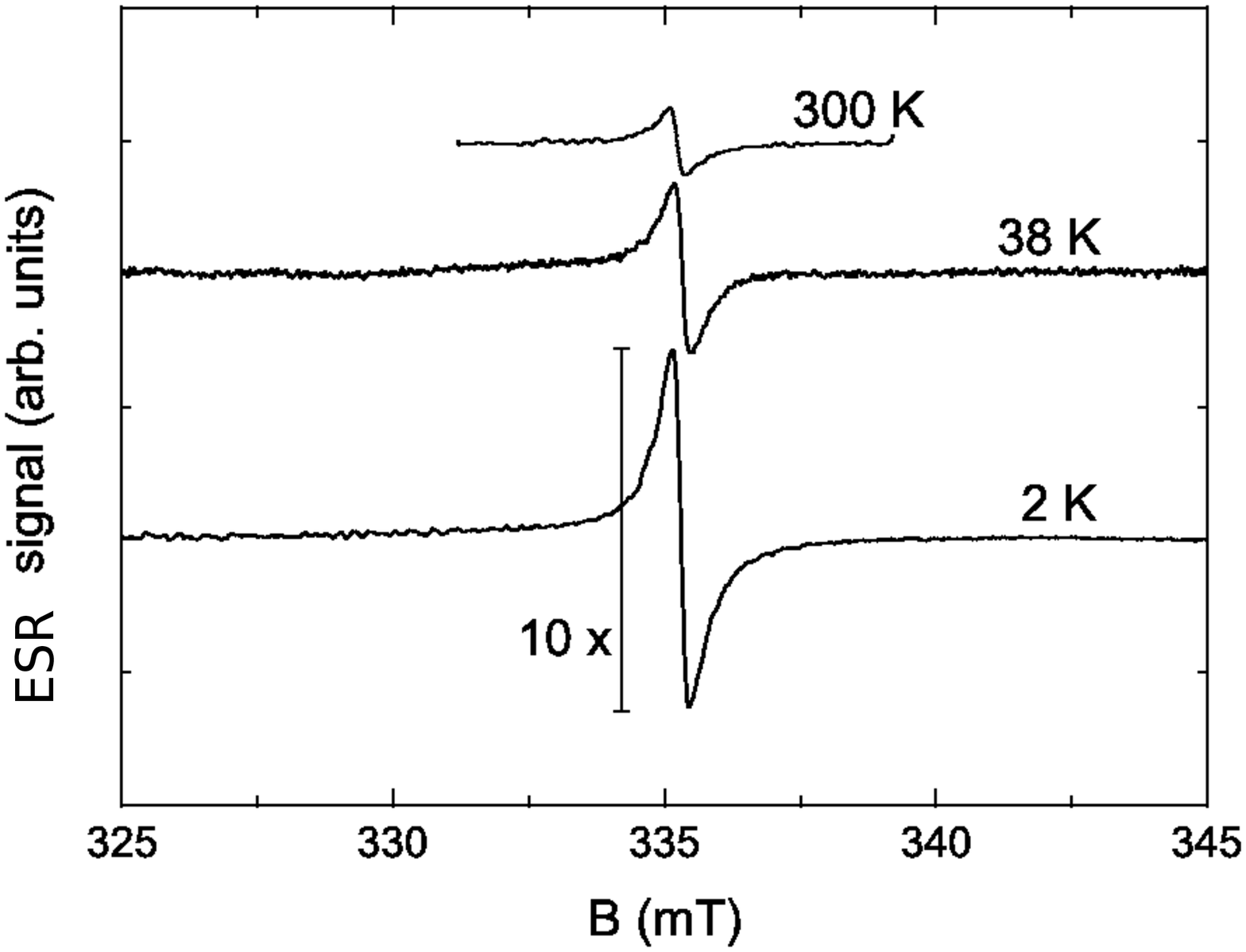}

\caption{\label{fig:ESR} Typical X-band ESR spectra of powdered sample measured
at $9.4\:\mathrm{GHz}$ at temperatures $T=2,\:38,\:\mathrm{and}\:300\:\mathrm{K}$.
The signal measured at $2\:\mathrm{K}$ is ten times reduced.}
\end{figure}

The X-band ESR spectra of carbon powder measured at temperatures $T=2,\:38,$
and $300\:\mathrm{K}$ are shown in Fig. \ref{fig:ESR}. The intensity
of the signal decreases significantly in the temperature range from
$2$ to $40\:\mathrm{K}$  following the temperature dependence of
the magnetization (Sec. \ref{sub:Magnetization-vs.-temperature}).
A slight asymmetry of the ESR line at temperatures above $40\:\mathrm{K}$
was observed, reminiscent of a metallic distortion due to the skin
effect, while at lower temperatures $T<40\:\mathrm{K}$ the line is
symmetric. We determined the $g$-factor to be $g=2.0026$ at $300\:\mathrm{K}$;
upon cooling down the $g$ factor changes to $g=2.0025$ at $2\:\mathrm{K}$,
both close to the free electron value. 

\subsection{\label{sub:Magnetization-vs.-temperature}Measurements of magnetization
vs temperature}

\begin{figure}
\includegraphics[width=6.5cm]{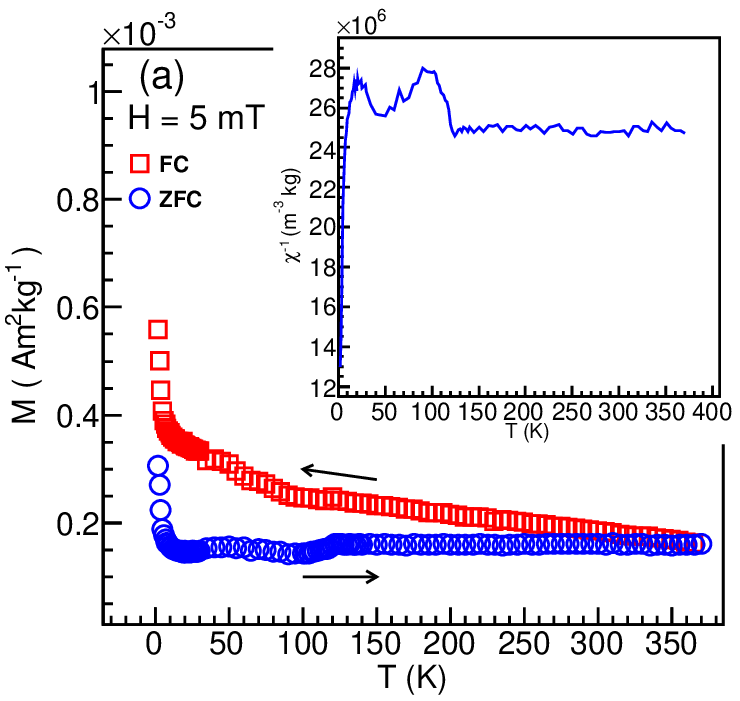}

\includegraphics[width=6.5cm]{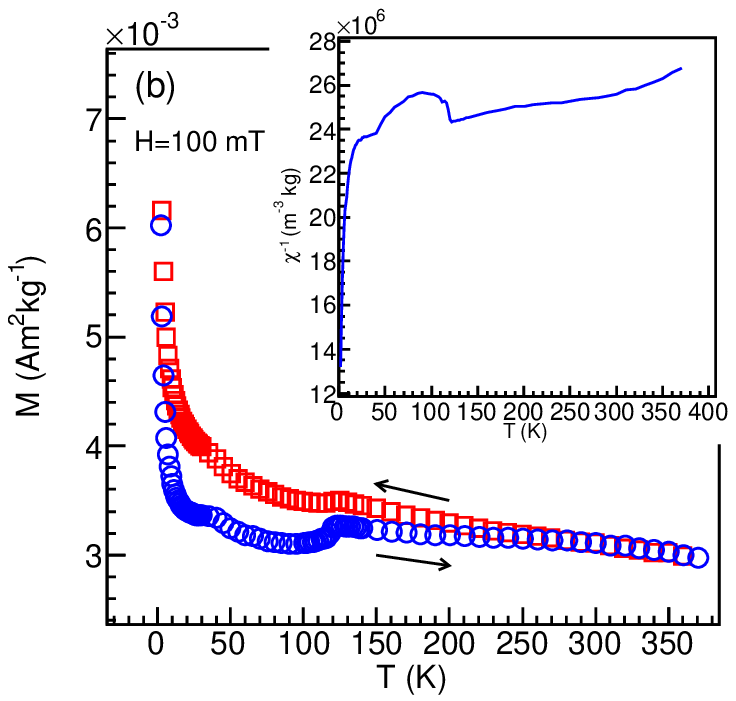}

\includegraphics[width=6.5cm]{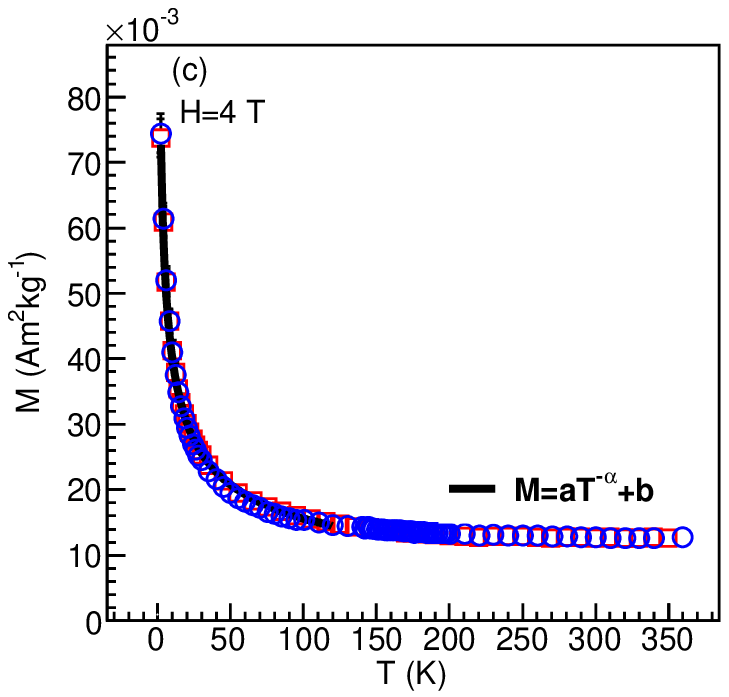}

\caption{\label{fig:MvsT}(Color online) Zero-field-cooled (ZFC) and Field-cooled
(FC) magnetizations at magnetic fields: (a) $H=5\:\mathrm{mT}$ with
diamagnetic magnetization correction $M_{D}=0\:\mathrm{Am^{2}kg^{-1}}$,
(b) $H=100\:\mathrm{mT}$ with diamagnetic correction $M_{D}=-0.0014\:\mathrm{Am^{2}kg^{-1}}$,
and (c) $H=4\:\mathrm{T}$ with diamagnetic correction $M_{D}=-0.056\:\mathrm{Am^{2}kg^{-1}}$.
The magnetization vs. temperature $M(T)$ is approximated by the function
$M(T)=aT^{-\alpha}+b$ where the exponent $\alpha=0.41\pm0.03$ was
found for temperature $2\leq T\leq100\:\mathrm{K}$ as shown by the
solid line.}
\end{figure}

The temperature dependence of the zero-field-cooled and field-cooled
magnetizations, $M_{ZFC}$ and $M_{FC}$ respectively, are shown in
Fig. \ref{fig:MvsT} for external magnetic fields $\mu_{0}H=0.005,\:0.1$,
and $4\mathrm{\: T}$ after subtraction of the diamagnetic background
$M_{D}$. For low magnetic fields $\mu_{0}H=0.005$ and $0.1\:\mathrm{T}$
{[}Figs. \ref{fig:MvsT}(a) and \ref{fig:MvsT}(b){]}, the magnetization
curves show irreversible behavior and a strongly increasing magnetization
for temperatures $2\leq T<20\:\mathrm{K}$. This irreversible property
tends to disappear when measurements are performed in higher magnetic
fields, for example $\mu_{0}H=4\:\mathrm{T}$ as in Fig. \ref{fig:MvsT}
(c). For magnetic fields $\mu_{0}H>10\:\mathrm{mT}$, the diamagnetic
signal begins to be comparable to the rest of the magnetization of
the sample, which results in a decrease in the total measured positive
magnetization signal, and at high fields the diamagnetism dominates
the measured sample response. Using a diamagnetic susceptibility $\chi_{D}=-17\times10^{-9}\mathrm{\: m^{3}kg^{-1}}$
(see next section), the diamagnetic magnetizations $M_{D}=-0.0014\:\mathrm{\; and}\:-0.056\:\mathrm{m^{2}kg^{-1}}$
were subtracted in order to get the net magnetizations from the measured
$M_{ZFC}$ and $M_{FC}$ curves for the magnetic fields $\mu_{0}H=0.1$
and $4\:\mathrm{T}$, respectively.

The insets of Figs. \ref{fig:MvsT}(a) and \ref{fig:MvsT}(b) show
the calculated inverse dc susceptibilities $\chi^{-1}=H/M$ vs temperature
$T$. For low magnetic fields $\mu_{0}H=0.005\:\mathrm{\; and}\:\;0.1\:\mathrm{\; T}$
and low temperatures $T<20\:\mathrm{K}$ {[}Fig. \ref{fig:MvsT} (a)
and \ref{fig:MvsT}(b){]}, the susceptibility does not show Curie-law-like
behavior $\chi\propto T^{-1}$. For the highest magnetic field $\mu_{0}H=4\:\mathrm{T}$
and temperature $2\leq T\leq100\:\mathrm{K}$, we found that the magnetization
curve $M(T)$ could be fitted to a temperature variable plus constant
parts which were approximated by a function $M(T)=aT^{-\alpha}+b$,
where $a=0.104\pm0.003\;\mathrm{Am^{2}kg^{-1}}$, $\alpha=0.41\pm0.03$,
and $b=9.8\times10^{-5}\;\mathrm{Am^{2}kg^{-1}}$ were determined
for magnetic field $\mu_{0}H=4\:\mathrm{T}$.

\subsection{Magnetization vs magnetic field}

Measurements of magnetization $M$ vs magnetic field $\mu_{0}H$ for
both increasing and decreasing field were performed to obtain the
full magnetization loops for several temperatures in the range $2\leq T\leq300\:\mathrm{K}$.

\begin{figure}
\includegraphics[width=7cm]{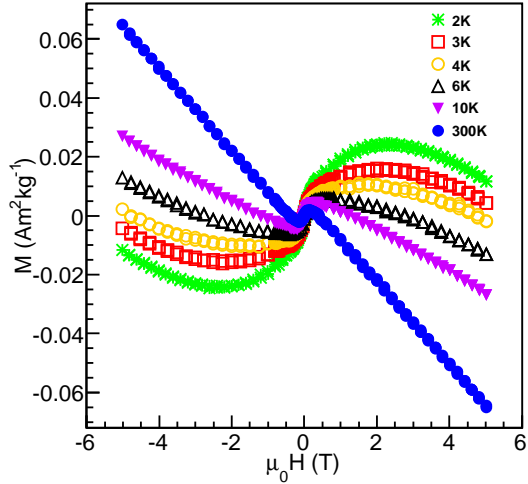}

\caption{\label{fig:Magnet_exp}(Color online) Magnetization $M_{E}$ vs magnetic
field $\mu_{0}H$ for temperatures $2\leq T\leq300\: K.$}
\end{figure}

The experimental isothermal magnetization loops are shown in Fig.
\ref{fig:Magnet_exp}. The total magnetization $M_{E}$ consists of
a contributions from the diamagnetic magnetization $M_{D}=-\chi_{D}H$,
a ferromagnetic part $M_{F}$, and an additional magnetization signal
$M$ with positive, paramagnetic sign. As the diamagnetic magnetization
$M_{D}$ varies linearly with the magnetic field and the ferromagnetic
magnetization $M_{F}(\mu_{0}H)$ is known from the mainly ferromagnetic
signal at ambient temperature, these components can be subtracted
in order to obtain the net magnetization $M=M_{E}-M_{D}-M_{F}$. The
diamagnetic contribution was determined as follows: A selection was
made of the linear parts of the magnetization curves $M_{E}$ vs magnetic
field $\mu_{0}H$ for the temperature $T=300\:\mathrm{K}$ and for
the strongest positive and negative magnetic fields, $4.5\leq\mu_{0}H\leq5.0\:\mathrm{T}$
and $-4.5\leq\mu_{0}H\leq-5.0\:\mathrm{T}$. The diamagnetic susceptibility
$\chi_{D}=-17\times10^{-9}\pm0.2\times10^{-9}\mathrm{m^{3}kg^{-1}}$
at the temperature $T=300\:\mathrm{K}$ was then determined by fitting
these two parts to one linear field behavior. 

\begin{figure}
\includegraphics[width=7.5cm]{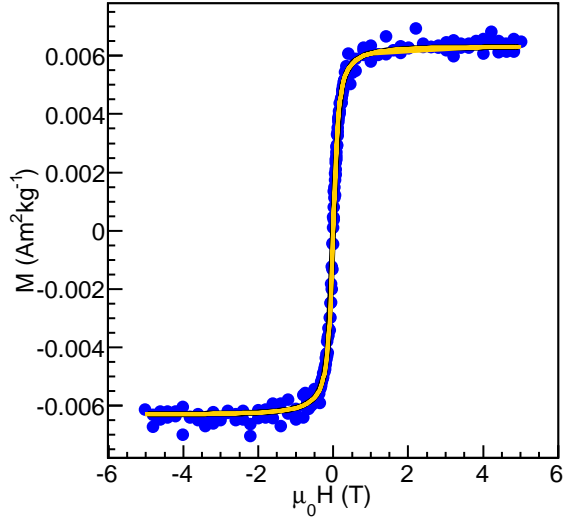}

\caption{\label{fig:MvzH_fer}(Color online) Magnetization $M_{F}$ vs magnetic
field $\mu_{0}H$ for temperature $T=300\:\mathrm{K}$. The experimental
data are approximated by the Brillouin function Eq. (\ref{eq:Bri}). }
\end{figure}

The ferromagnetic contribution $M_{F}$ to the total magnetization
$M_{E}$ is approximately independent of the temperature in the range
$15\leq T\leq300\:\mathrm{K}$. The experimental isothermal magnetization
loop $M_{F}$ at $T=300\:\mathrm{K}$ was well modeled by the Brillouin
function Eq. (\ref{eq:Bri}) where the free fitting parameters were
$T=1.34\:\mathrm{K}$, $g=2.0$, $S=17$, and $M_{S}=6.3\times10^{-3}\:\mathrm{A}\mathrm{m}^{2}\mathrm{kg^{-1}}$
as shown in Fig. \ref{fig:MvzH_fer}. Coercivity fields $H_{C}=20$
and $10\mathrm{\: mT}$ were found at temperatures $2$ and $300\:\mathrm{K}$,
respectively. This ferromagnetic contribution is consistent with the
Fe impurity level found in the samples (see Sec. \ref{sec:Conclusions}). 

\begin{figure}
\includegraphics[width=7.5cm]{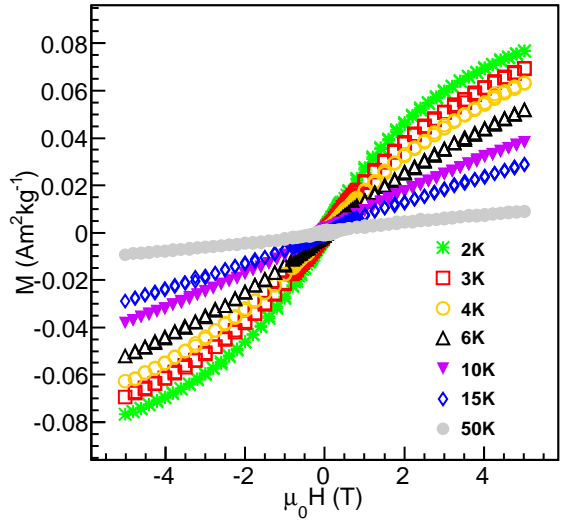}

\caption{\label{fig:net_mag}(Color online) Magnetization $M$ vs temperature
$T$ after correction of diamagnetic ($\chi_{D}=-17\times10^{-9}\mathrm{m^{3}kg^{-1}}$)
and ferromagnetic magnetization $M_{F,\, T=300\,\mathrm{K}}$ contributions. }
\end{figure}

The diamagnetic magnetizations $M_{D}=-\chi_{D}H$ and isothermal
magnetization $M_{F,\, T=300\,\mathrm{K}}$ were subtracted from the
experimentally measured magnetization values $M_{E}$ in Fig. \ref{fig:Magnet_exp}.
The resultant magnetization curves $M(\mu_{0}H)$ are shown in Fig.
\ref{fig:net_mag}.

The field dependence of the magnetization in Fig. \ref{fig:net_mag}
may resemble the magnetization of free, noninteracting spins. The
paramagnetic magnetization of a free-spin system is described by the
Brillouin function and is based on the assumption that the population
of energy levels obeys Boltzmann statistics \cite{Stanley}. The rescaled
magnetization, i.e., the magnetization relative to its saturation
value $M_{S}$, is then $M/M_{S}=B_{S}(x)$. Here, the Brillouin function
$B_{S}(x)$ is given by

\begin{equation}
B_{S}(x)=\frac{2S+1}{2S}\coth\left(\frac{2S+1}{2S}x\right)-\frac{1}{2S}\coth\left(\frac{1}{2S}x\right)\label{eq:Bri}
\end{equation}
 with $x=g\mu_{B}\mu_{0}HS/k_{B}T$, where $S$ is the spin value,
$g$ is the Land\'{e} factor, $\mu_{B}$ is the Bohr magneton, $\mu_{0}$
is the permeability of vacuum, and $k_{B}$ is the Boltzmann constant.
The saturation magnetization is $M_{S}=Ng\mu_{B}S$, where $N$ is
the number of magnetic moments per unit volume. For the specific case
$S=1/2$ Eq. (\ref{eq:Bri}) transforms into the hyperbolic tangent,
$B_{1/2}(x)=\tanh(x).$

\begin{figure}
\includegraphics[width=7.5cm]{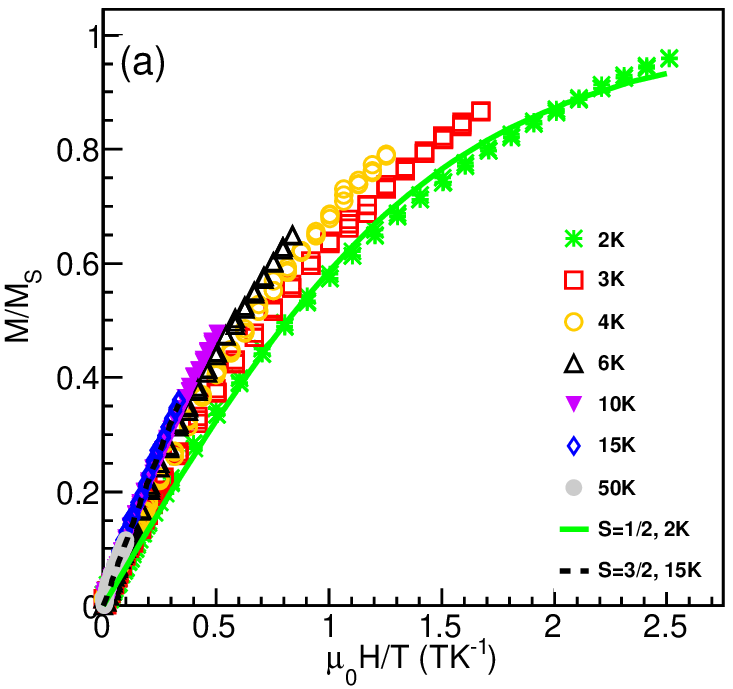} \includegraphics[width=7.5cm]{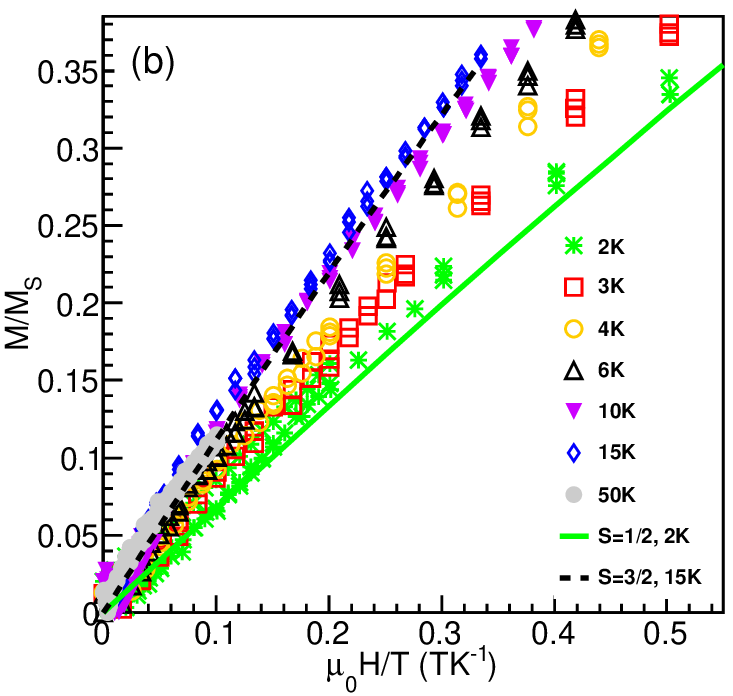}

\caption{\label{fig:replotMvsH}(Color online) (a) $M/M_{S}$ vs $\mu_{0}H/T$
plots for low temperatures $2\leq T\leq50\:\mathrm{K}$. The data
are fitted to Brillouin functions Eq. (\ref{eq:Bri}) with parameters
$g=2.0$, $S=\frac{1}{2}$ for $T=2\:\mathrm{K}$ (solid line), and
$g=2.0$, $S=\frac{3}{2}$ for $T=15\:\mathrm{K}$ (dashed line).
(b) Magnification of the lower left-hand corner of (a).}
\end{figure}

Figure \ref{fig:replotMvsH} shows the rescaled magnetization data
$M/M_{S}$ replotted as a function of $\mu_{0}H/T$, keeping $M_{S}$
fixed at the value found at 2K. The data for $T=2\:\mathrm{K}$ were
relatively well modelled by a localized-electron, free-spin model
with $g\doteq2$ and $S=\frac{1}{2}$, plotted as a green solid curve
in Fig. \ref{fig:replotMvsH}. For temperatures $15\leq T\leq50\:\mathrm{K}$,
fits with $S=\frac{1}{2}$ showed poor agreement with the data. Much
better fits were obtained by a Brillouin function with the higher
spin value $S=\frac{3}{2}$. As seen in the figure, for temperatures
$2<T<15\:\mathrm{K}$ the graphs do not collapse into a single function,
which is a typical signature of a free, localized spin model. Apparently,
it is not possible to apply such a description  for these temperatures.

\section{\label{sec:Discussion}Discussion}

The magnetization of the present carbon powder can be influenced by
several factors. For example, particle shape (tube, disk, or cone),
level of defects within the particles, density of magnetic impurities
such as iron, and presence of gas molecules adsorbed in the sample.
The contributions of these factors to the total sample magnetization
$M$ can be diamagnetic, paramagnetic, or ferromagnetic in specific
temperature intervals. The amount of iron impurities and their distribution
in the carbon samples are thus important attributes to evaluate due
to their great impact on the magnetic properties of the samples. \cite{Esquinazi_2007,Esquinazi_2010}
If the size of the magnetic impurity particles is big enough, then
such particles can behave ferromagnetically and give rise to a ferromagnetic
contribution to the sample magnetization. Small magnetic impurities
uniformly distributed in the sample can behave as non-interacting
magnetic moments which show paramagnetic behavior. For example, pure
iron ($\mathrm{Fe}$) or magnetite ($\mathrm{Fe_{3}O_{4}}$) of impurity
density level $1\:\mu\mathrm{g/g}$ in the form of big particles can
contribute to the saturation magnetization $M_{S}$ amounting to $2.2\times10^{-4}\:\mathrm{Am^{2}}\mathrm{kg^{-1}}$
or $1.0\times10^{-4}\mathrm{\: Am^{2}kg^{-1}}$, respectively, at
room temperature. \cite{Hohne_2008} On the other hand, if the magnetic
impurities are small and uniformly distributed their paramagnetic
contribution is much smaller. Taking into account the average density
of iron impurities $60\:\mathrm{\mu g}/\mathrm{g}$ (Sec. \ref{sub:Impurities-measurements})
and assuming that the iron particles are big enough and behave ferromagnetically,
then their contribution to the saturation magnetization $M_{S}$ at
room temperatures is expected to be about $13\times10^{-3}\mathrm{\: Am^{2}kg^{-1}}$
for iron impurities and $6\times10^{-3}\mathrm{\: Am^{2}kg^{-1}}$
if they are magnetite impurities. Magnetization measurements showed
a ferromagnetic contribution with saturation magnetization $M_{S}=6.3\times10^{-3}\:\mathrm{Am^{2}kg^{-1}}$
(Fig. \ref{fig:MvzH_fer}), which falls inside this range of theoretical
saturation values. However, ZFC and FC measurements (Fig. \ref{fig:MvsT})
do not show clear evidence of a blocking temperature associated with
an Fe ferromagnetic phase. It is highly probable that the ferromagnetic
part of the signal at all temperatures is the result of iron-containing
microparticles, which could stem from the pyrolitic production process.
\cite{Nair} However, no such particles have been identified in electron
microscopy images or in energy-dispersive x-ray (EDX) spectra. We
conclude that in the present samples the expected contribution to
the saturation magnetization $M_{S}$ from magnetic iron impurities
is comparable to the measured magnetization at room temperatures.
However, at low temperatures $T<100\:\mathrm{K}$ the dominating contribution
to $M_{S}$ comes from the carbon particles. Thus, the isothermal
magnetization function $M_{F,\, T=300\,\mathrm{K}}$ at temperature
$T=300\:\mathrm{K}$ was subtracted from the experimental magnetization
curves to correct for the ferromagnetic contribution from the magnetic
impurities as explained in Sec. III C.

The total magnetization of the samples is composed of a relatively
strong negative diamagnetic signal and a smaller positive magnetization.
This combination is a common feature of the magnetization of HOPG
samples. \cite{Hohne_2008,Cervenka,Nair} It differs from the behavior
for nanodiamond powders \cite{Levin} and carbon nanofoams, \cite{Rode}
where the magnetization is dominated by the paramagnetic contribution
of orbital electrons. The diamagnetic susceptibility of the carbon
powder is similar to what has been observed for other carbon allotropes,
for example the HOPG samples \cite{Cervenka,Hohne_2008} and mono-
and bilayer graphene crystallites of sizes $10$ to $50\;\mathrm{n}\mathrm{m}$.
\cite{Sepioni} This common feature is explained by delocalized $\pi$
electrons in carbon rings where currents are induced by the external
magnetic field. \cite{Haddon} The magnetic susceptibility of diamond
and $\mathrm{C}_{60}$ does not depend strongly on the temperature.\cite{Haddon}
On the other hand, the diamagnetic susceptibility of a 2D honeycomb
carbon lattice has been calculated to be temperature dependent with
an absolute value that increases with temperature. \cite{Castro_Neto,Haddon}
In order to simplify the separation of diamagnetic, paramagnetic,
and ferromagnetic contributions we had to assume that the diamagnetic
contribution is constant and independent of temperature. 

For low magnetic fields {[}Figs. \ref{fig:MvsT} (a) and (b){]}, the
temperature dependences of ZFC and FC magnetizations are irreversible
in a wide temperature range. This observation agrees with previous
results showing thermal hysteresis. \cite{Mombru,Arcon,Kopelevich,Boukhvalov}
It is often assumed that such magnetization behavior originates from
isolated spin clusters \cite{Arcon,Kopelevich,Boukhvalov} which could
display a spin-glass like state. \cite{Arcon} As seen in Figs. \ref{fig:MvsT}
(b) and \ref{fig:MvsT}(c), for temperatures $2\leq T<100\:\mathrm{K}$
and magnetic fields $0.1\leq\mu_{0}H\leq4\:\mathrm{T}$, the magnetization
follows a power law $M\sim T^{-\alpha}$ with exponent $\alpha<1$.
The exponent $\alpha<1$ differs from that of nanodiamond powder,
\cite{Levin} graphene sheets, \cite{Sepioni} and HOPG samples, \cite{Nair}
where a Curie law behavior, $\alpha=1$, was observed for magnetic
fields of $1\:\mathrm{T}$. It has been found earlier that certain
magnetic materials, for example doped semiconductors \cite{Rode}
or certain rare-earth intermetallics, \cite{Ghost} show exponents
$\alpha<1$ for low temperatures and low magnetic fields. An exponent
$\alpha<1$ indicates that there are magnetic spin interactions. \cite{Ghost}
Bhatt and Lee \cite{Bhatt} found an exponent $\alpha<1$ for a 3D
model of spatially random Heisenberg spins $S=\frac{1}{2}$ that interact
through an exponentially decaying interaction vs separation. The ESR
spectra in Fig. \ref{fig:ESR} show a tendency of localization of
electrons when the temperature decreases ($T<38\:\mathrm{K}$). For
the lowest temperature $T=2\:\mathrm{K}$, the ESR spectrum resembles
the spectrum of an insulator and the magnetization vs magnetic field
is well approximated by the Brillouin model of noninteracting spins
$S=\frac{1}{2}$. For temperatures $2<T<15\:\mathrm{K}$, the magnetization
is higher than that predicted by the Brillouin function-based model
for $S=\frac{1}{2}$. However, at $T=15\:\mathrm{K}$ it is well approximated
by a Brillouin function with $S=\frac{3}{2}$. It is possible that
both localized and itinerant magnetic processes can coexist in this
temperature range, resulting in a behavior that looks like a smooth
change of the apparent average spin value $\langle S\rangle$ from
$\frac{1}{2}$ to $\frac{3}{2}$ as temperature increases (Fig. \ref{fig:replotMvsH}).
Similarly, for $15\leq T<50\:\mathrm{K}$ the magnetization can be
consistently approximated by a Brillouin function with spin $S=\frac{3}{2}$
(data for $T=20$, 30, and 40 K are not shown). The magnetization
vs. temperature results are similar to the results observed for carbon
nanofoams. \cite{Rode} Contrary to our observation of a changing
$S$ value, it has been found that for graphene sheets \cite{Sepioni}
$S=2$ and $S=\frac{5}{2}$ and for graphene sheets with induced point
defects \cite{Nair} $S=\frac{1}{2}$, independent of temperature.
One possible explanation for the differences between our results and
these results could be the different structure of the graphene sheets
and the present nanoparticles, which consist of a mixture of crystalline
and disordered phases where each phase can contribute separately to
the sample magnetization. The behavior of the spins at the interface
between these phases is unknown. To clarify these points, future studies
are needed of samples containing purified phases.

\section{\label{sec:Conclusions}Conclusion}

We have measured magnetization properties of carbon powder samples
containing carbon cones and disks. A ferromagnetic contribution is
consistent with the known amount of Fe impurities in the sample and
was identified and subtracted from the data. The measurements showed
thermal hysteresis in the magnetization for weak fields which we believe
is due to magnetic ordering intrinsic to the carbon particles. Based
on the results of ESR spectra and magnetization vs temperature and
magnetization vs magnetic field measurements, it seems that there
exist localized electrons at temperature $T=2\:\mathrm{K}$, and their
response to magnetic field is well described by a free-spin $S=\frac{1}{2}$
model. For temperatures $2<T<50\:\mathrm{K}$ the magnetization is
higher than that calculated from the Brillouin function for this model,
which may be a result of interactions among localized- or intinerant-
electron spins in this temperature range. Similar magnetic behavior
has been found in other carbon-based materials like HOPG, diamond,
nanofoams, and graphene sheets, but there are also clear differences
between the current material and the earlier reports on other samples.
More studies are needed to learn more about the complicated interactions
among localized and itinerant magnetic moments in the growing class
of carbon nanomaterials.
\begin{acknowledgments}
The authors would like to thank Jaroslav Kr\'{a}l for valuable discussion
and Jaroslav \v{C}ern\'{y} for operating the Van de Graaff accelerator
as well as J. P. Pinheiro of n-Tec AS for providing the samples used
in this study. This work was supported by Institutional Research Plan
No. MSM6840770040, by SAS Centre of Excellence CFNT MVEP, and by Research
Council of Norway Grant No. 191621/F20.\end{acknowledgments}

\end{document}